\documentclass[slac_one]{revtex4}
\usepackage{graphicx}
\usepackage{fancyhdr}
\begin{document}

\title{DMTPC-10L: Direction-Sensitive Dark Matter Detector Prototype}

\author{D. Dujmic\footnote{Corresponding author (ddujmic@mit.edu)}, 
P. Fisher, S. Henderson, R. Lanza, J. Lopez, 
A. Kaboth, G. Kohse, J. Monroe, R. Vanderspek, G.Sciolla, R. Yamamoto  }
\affiliation{Massachusetts Institute of Technology, Cambridge, MA 02421, USA }

\author{S. Ahlen, K. Otis, A. Roccaro, H. Tomita}
\affiliation{Boston University, Boston, MA 02215, USA}

\author{N. Skvorodnev, H. Wellenstein}
\affiliation{Brandeis University, Waltham, MA 02454, USA }

\begin{abstract}
The known direction of  motion of dark matter particles relative to the Earth
may be a key for their unambiguous identification even in the presence of backgrounds.
We describe a prototype detector that is able to  reconstruct direction 
vectors of weakly interacting massive particles that may the 
dominant constituent of the dark matter in our galaxy. 
The detector uses a low-density gas (CF$_4$) in a 10liter time-projection chamber 
with mesh-based electrodes and optical and charge readout. 
Initial results confirm good performance in the reconstruction of direction angle and 
sense ('head-tail') for low-momentum nuclear recoils.
\end{abstract}

\maketitle

\thispagestyle{fancy}

\section{INTRODUCTION}

Observation of weakly interacting massive particles (WIMPs) may require  a good background rejection
and a correlation of the WIMP direction with the galactic motion through the dark matter halo.
Both requirements can be accomplished with detectors using low-pressure gas as the target material.
However, the low density of  gaseous detectors requires large detector volumes, 
and the shortness of signal tracks necessitate  fine detector granularities.
In the case of CF$_4$ as the detector material, one ton of gas occupies a volume of approximately 
$16\times16\times16$~m$^3$ at 50~Torr of pressure, and 
a 50~keV fluorine recoil travels 1.5~mm.
Therefore, a multi-cubic meter detector with resolution of the order of hundreds of micrometers 
is needed for a directional dark matter experiment.

\section{DMTPC-10L DETECTOR}

%\subsection{Detector design}

\begin{figure*}[t]
\centering
\includegraphics[width=14cm]{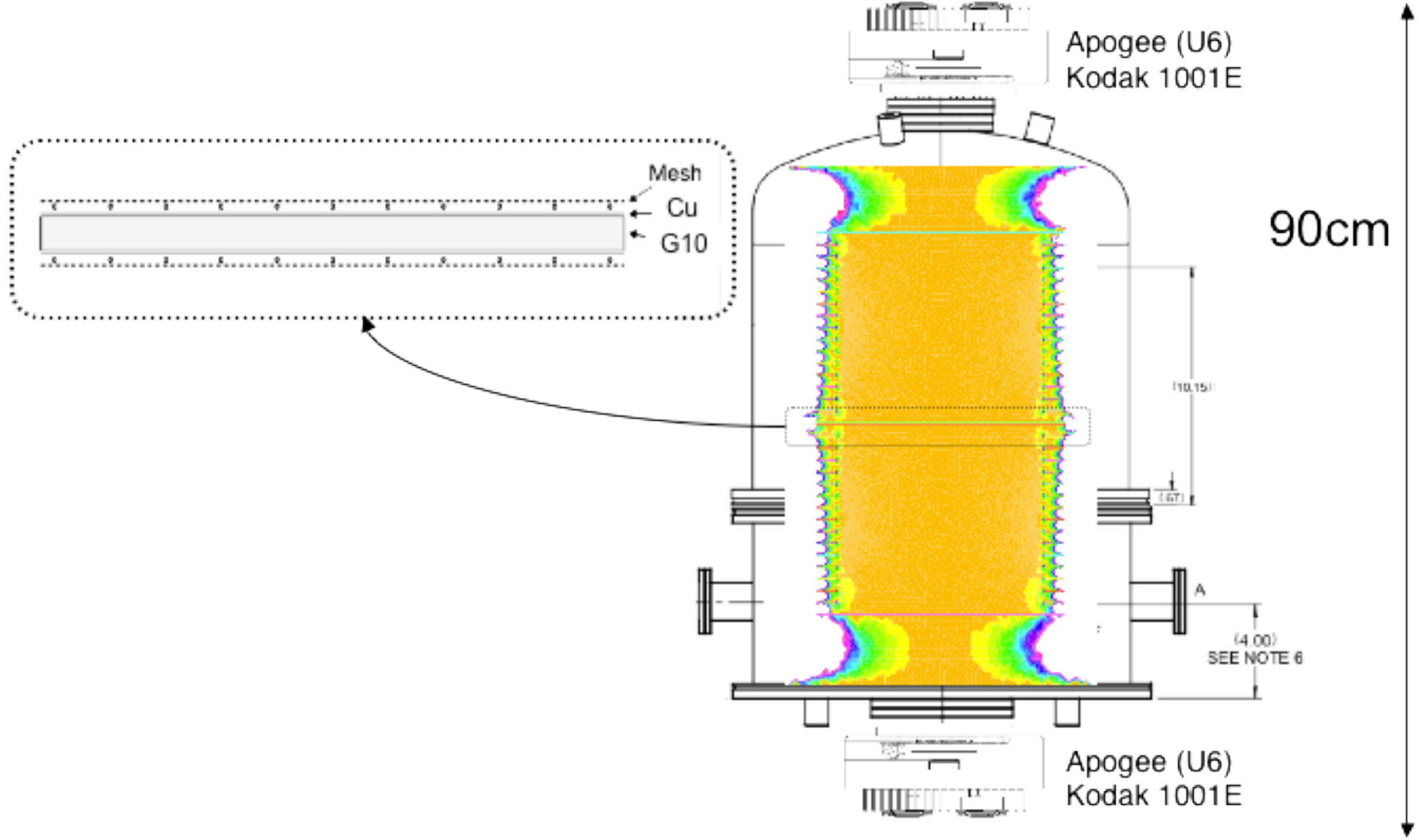}
\caption{Contours of a vacuum vessel holding a detector with 10-liter fiducial volume.  
The drift cage consists of stainless steel rings with 25~cm inner diameter.
A calculation of field uniformity ($|E_\perp|/E_{tot}$) in the drift cage is superimposed
to the drawing, with each color shade corresponding to 1\% of the change in the normalized transverse field.
The enlarged section of the amplification is shown on the left, with the field created in between a grounded mesh and
a solid copper anode separated by 0.5~mm fluorocarbon wires. Two identical structures are used for the two
drift regions. The scintillation light is recorded by two CCD cameras on top and bottom, and charge is read out of the two anodes. } \label{vessel}
\end{figure*}

A schematic of the DMTPC detector is shown in Figure~\ref{vessel}.
A stainless steel vessel is filled with 50-100~Torr of CF$_4$ gas and holds two back-to-back time-projection-chambers (TPC's).
Ionization electrons that are created by recoiling nucleus 
are drifted in an electric field ($\sim 400$~V/cm) toward the amplification region so they can be detected.
Each drift cage is made of stainless steel rings with the inner diameter of 25~cm and a total height of 25~cm.
The separation and the width of the rings are optimized using a calculation based on the finite elements method
in order to achieve  uniform electric field in the vertical direction with a minimum number of drift rings.
The perpendicular (radial) component of the electric field is less than 1\% in the entire fiducial volume.
The height of the drift region is limited by the electron diffusion~\cite{Dujmic:2007bd}.

A double-sided amplification region uses a solid copper for the anode and a stainless steel
mesh with 256~$\mu$m pitch as the grounded electrode. A detailed description and performance evaluation of such amplification system is
given in Ref.~\cite{Dujmic:2008ut}.
Scintillation photons created during the avalanche charge multiplication
are collected by  Nikon photographic lens with f-stop ratio of  1.2 and focal distance of 55~mm, and 
are recorded with Apogee U6 cameras using a Kodak 1001E CCD chip.
The total area imaged by each CCD is $16 \times 16$ cm$^2$ so the total active volume is approximately 10 liters.

The avalanche creates approximately 3 electrons for each scintillation photon with the wavelength between 200-800~nm,
or approximately an order of magnitude more electrons than photons in the wavelength range that is accessible to 
CCD detectors~\cite{Kaboth:2008mi}.
We use the charge readout to improve the energy resolution of low-momentum nuclear recoils. The signal is collected from
the anode and passed through a charge-sensitive preamplifier  (Ortec PC109) and shaping amplifier (Ortec 485) before being recorded with
a digitizer (Tek 1002).

\section{CALIBRATION RESULTS}

\begin{figure*}[t]
\centering
\includegraphics[width=7cm]{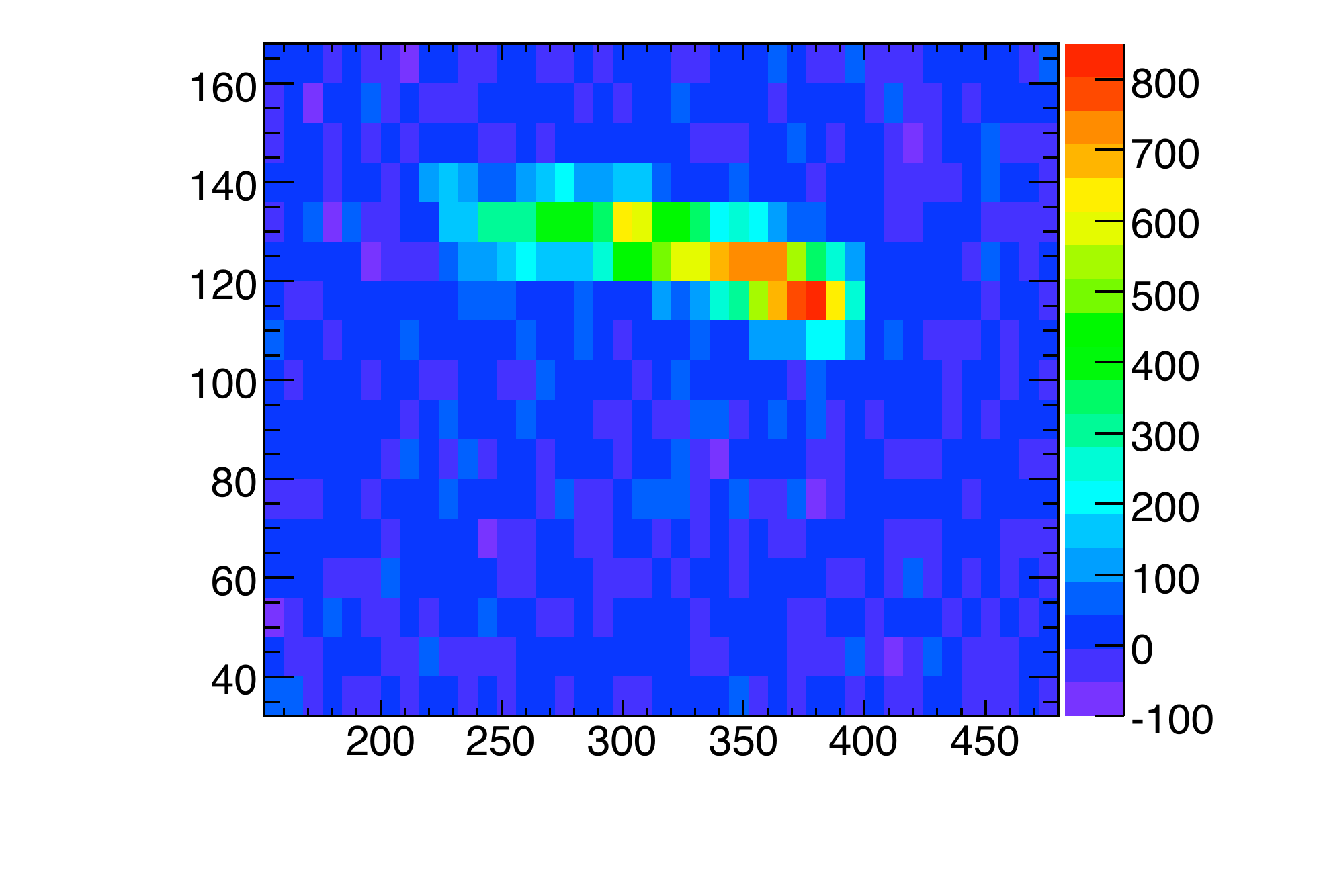}
\includegraphics[width=7cm]{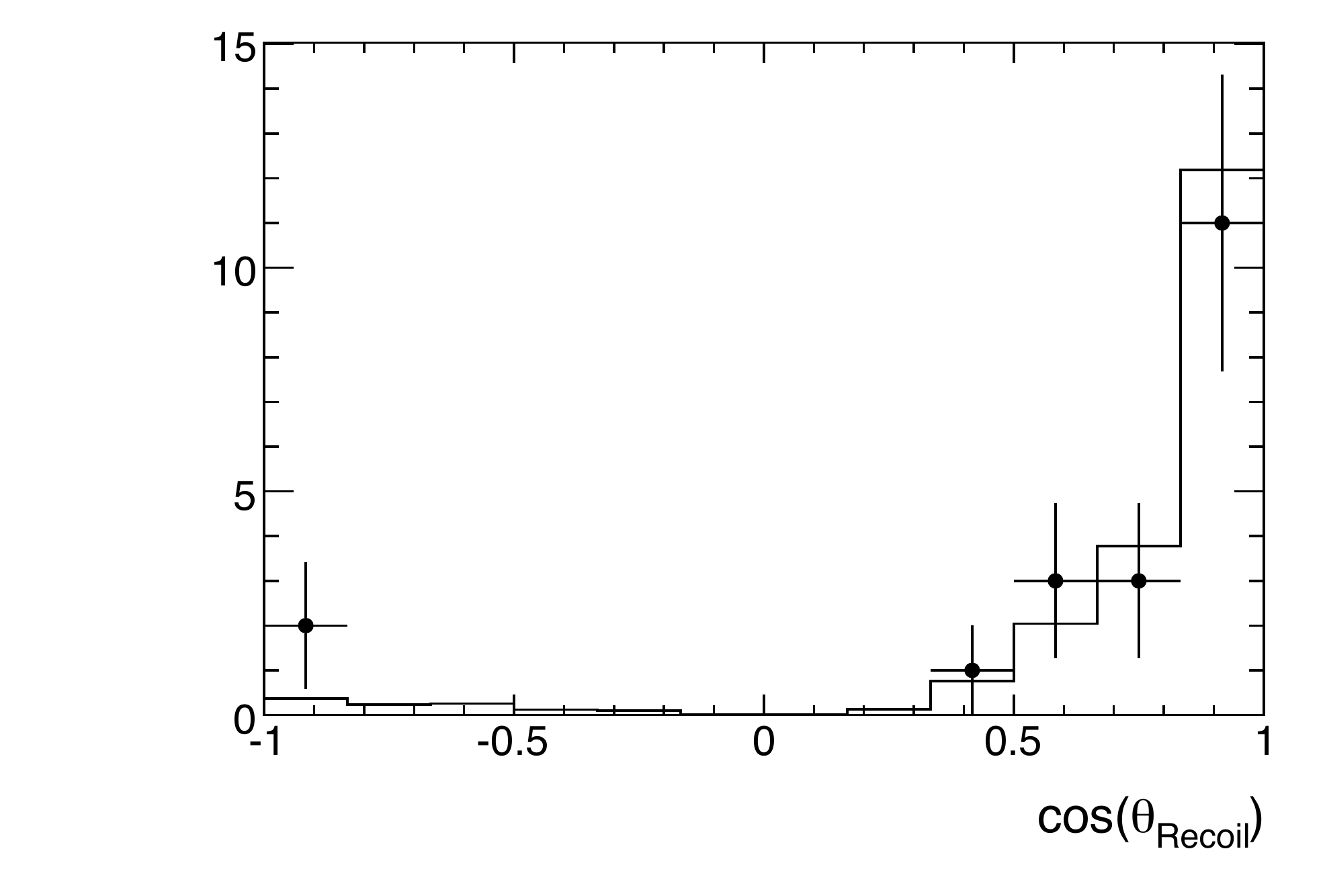}
\caption{Detector response: A CCD image of nuclear recoil in exposure to neutrons from
a $^{252}$Cf source (left), and a distribution of reconstructed angles with head-tail information in
data and simulated samples (right).} \label{fg::recoil}
\end{figure*}

We demonstrate the ability to determine the directional sense of low-momentum recoils, the `head-tail' effect,
by exposing the detectors to neutrons from a $^{252}$Cf source.
The ionization rate of a low-momentum  nuclear recoil decreases as it slows down in the detector gas,
and the sense of its direction can be deduced from the scintillation profile along the track.
An image of the recoil track is shown in the left plot of  Figure~\ref{fg::recoil}.
Neutrons are incident from the right, and the nuclear recoil also propagates from the right, 
which is evident from a decreasing light intensity.
The sense of  direction in single events can be reconstructed down to approximately 100~keV of recoil energy~\cite{Dujmic:2007bd},
but a larger data sample is needed to determine the full potential of the detector.

\begin{figure*}[t]
\centering
\includegraphics[width=7cm]{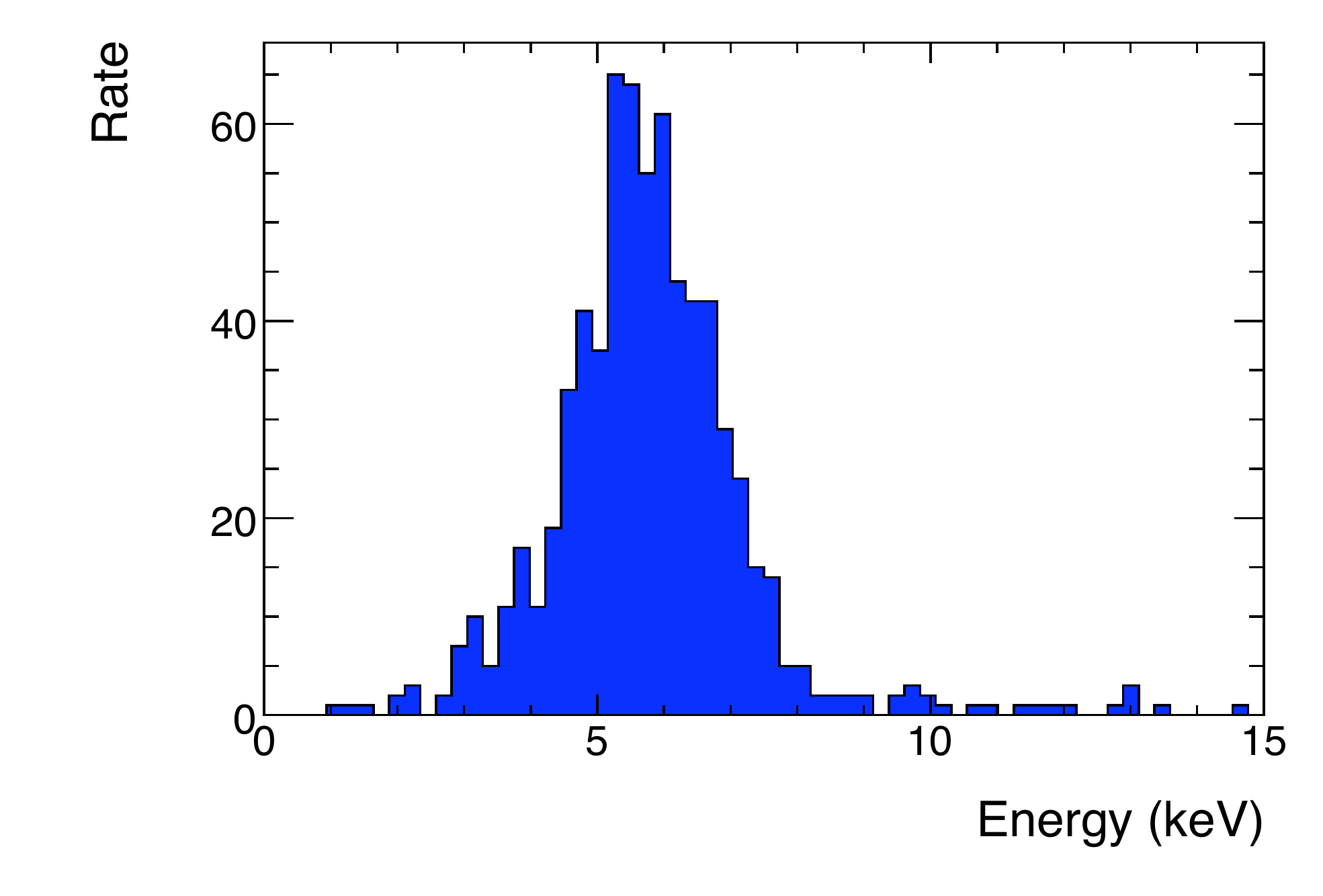}
\caption{A histogram of charge peak heights in exposure to 5.9 x-rays from
$^{55}$Fe source.} \label{fg::xray}
\end{figure*}

In order to measure the energy resolution, we expose the detector to x-rays from $^{55}$Fe source.
A histogram of amplitudes of signal pulses is shown in the right plot in Figure~\ref{fg::xray}.
The peak corresponds to 5.9~keV x-rays from K-$\alpha$ lines, and its width gives the energy resolution of approximately 19\%.

\section{SUMMARY AND PLANS}

Directional detection of dark matter  requires large  detector volumes with fine granularities.
We have built a prototype detector consisting of two TPC modules filled with low-pressure CF$_4$ gas,  
and demonstrated reconstruction of the directionality and directional sense. 
A detector system based on this technology can improve   current  limits on spin-dependent dark matter interactions
with approximately $0.1~\rm{kg}\cdot\rm{y}$ of CF$_4$ exposure, 
and test New Physics models with approximately $100~\rm{kg}\cdot\rm{y}$~\cite{Sciolla:2008rn}. 
We are currently taking data with the 10-liter prototype at the surface laboratory,
 and making preparations for operation in an underground laboratory.
Completion of these tests will allows us to design a cubic-meter module that will be a 
 building block of a ton-scale detector.

\section{ACKNOWLEDGMENTS} \label{Ack}

We wish to thank Hayk Yegorian for his valuable help in electric field calculation 
and charge readout tests, and Timur Sahin for creation of GEANT4 simulation of the detector.

We  also wish to thank the Office of Environment, Health \& Safety 
and the Laboratory for Nuclear Science MIT for technical support.
We acknowledge support by the Advanced Detector Research Program of 
the U.S. Department of Energy (contract number 6916448), 
the Reed Award Program,  the Ferry Fund, the Pappalardo Fellowship program, 
the MIT Kavli Institute for Astrophysics and Space Research,
and the Physics Department at the MIT.

\end{document}